\documentclass[12pt]{iopart}
\usepackage{iopams}  
\usepackage{graphicx}
\begin{document}

\title[Preserved entropy and fragile magnetism]{Preserved entropy and fragile magnetism}

\author{Paul C. Canfield and Sergey L. Bud'ko}
\address{Ames Laboratory US DOE and Department of Physics and Astronomy, Iowa State University, Ames, Iowa 50011, USA}

\vspace{10pt}

\begin{abstract}
A large swath of strongly correlated electron systems can be associated with the phenomena of preserved entropy and fragile magnetism.  In this overview we present our thoughts and plans for the discovery and development of lanthanide and transition metal based, strongly correlated systems that are revealed by suppressed, fragile magnetism or grow out of preserved entropy.  We will present and discuss current examples such as YbBiPt, YbAgGe, YbFe$_2$Zn$_{20}$, PrAg$_2$In, BaFe$_2$As$_2$, CaFe$_2$As$_2$, LaCrSb$_3$ and LaCrGe$_3$ as part of our motivation and to provide illustrative examples. 
\end{abstract}

%
%
\submitto{\RPP}
%
\maketitle
%
%

A goal of condensed matter physics is to map Hamiltonians onto compositional and structural phase space.  This goal lies somewhere between laudable and quixotic, but it is becoming increasingly tantalizing in its apparent proximity.  For simple systems, $s-p$ shell elements and compounds, many structures and properties can be understood and even predicted in this manner.  (Although it could be argued that MgB$_2$ \cite{nag01a}, a purely $s-p$ shell material, gives even this simple statement the lie.)  Strongly correlated systems, on the other hand, are an extreme example of this goal; they often come with competing energy scales that (i) make computational, post-priori, explanations difficult and (ii) render a-priori predictions of properties or ground states almost impossible.  

Strongly correlated systems, in this sense, can be thought of in literary terms as well.  In a play, the pleasure is to watch the characters evolve and interact, with each other and their setting.  Interesting plays often involve comparably strong or compelling characters striking sparks off of each other as the play progresses. "Who's afraid of Virginia Wolfe", "Cat on a Hot Tin Roof", "Taming of the Shrew" and "Lysistrata", are stark examples of this dynamic.  In a strongly correlated system there are often competing groundstates and / or comparable energy scales that spar with each other as the sample is cooled or perturbed.  Whereas copper and silicon are very useful and understanding them was one of the early successes of solid state physics, they do not offer examples of such drama.  They have one salient energy scale and one dominant character wandering the stage alone.  It is no wonder that computationally these are less difficult to master.

Although thinking about correlated electron systems in such literary terms helps to explain why they are hard to predict, it does not help find them, or even motivate the search for them.  Another way of thinking about many of the correlated systems of current interest, ones associated with real or imagined proximity to quantum phase transitions or quantum critical points, is to consider the preservation of entropy to low temperatures.  In many systems, following the entropy is as essential as “following the money” in many political situations.  Whereas a cooling through phase transition is associated with a removal of entropy, the suppression of the same phase transition can be associated with a postponement or reduction of the change in entropy.  If low enough temperatures can be reached with enough preserved entropy, then new, sometimes exotic, sometimes unexpected states can emerge.  

In this paper we will outline our current thinking about how preserved magnetic entropy and proximity to fragile magnetic states can be used as a way of rationalizing the search for new correlated electron systems.  We will do this in three parts, first with Yb-based compounds, i.e. systems with a hybridizing rare earth that is also a Kramers ion, and therefore have an aliquot of guaranteed magnetism and entropy.  Next, as a transitional example, Pr-based compounds will be discussed.  Since Pr is a non-Kramers ion there is not a finite, minimum quantity of entropy associated with the crystal electric field (CEF) splitting of the Hund's rule  ground state.  As a result, some care must be taken to preserve entropy to low enough temperatures.  Finally, we will end with a discussion of fragile magnetism in transition metal based intermetallic compounds.  In this case a great deal of effort must be spent on identifying systems that can actually support some degree of magnetism, i.e. for many intermetallic systems containing transition metals, the transition metal drops all hints of magnetism, local or itinerant, and for all intents and purposes behaves much like metallic copper or magnesium.  The progression, then, will be from systems with a more or less assured source of magnetic entropy as well as more or less built in fragility to systems that require increasingly more time and effort to have even some, hopefully fragile, magnetism.
\\

Yb-based compounds, much like Ce-based compounds, contain a hybridizing rare earth ion that is also a Kramers ion when it is in the trivalent state.  This means that, for heavy fermion (HF) compounds, there is a guaranteed R$ln2$, minimum entropy to be removed at low temperatures.  As will be discussed in detail below, this guaranteed minimum entropy is a luxury when discussing possible quantum critical states.  Whereas a relatively large number of Ce-based heavy fermion compounds were known as of 1990 \cite{ste84a,bra84a}, the discovery of Yb-based compounds lagged, due, in large part, to the high vapor pressure of Yb making arc-melting of polycrystalline samples problematic if not impossible.  Over the past 25 years, though, solution growth of Yb-based materials has led to the discovery of a diverse set of Yb-based heavy fermion materials.  Some of the better studied systems have been YbBiPt \cite{can91a,fis91a}, YbNi$_2$B$_2$C \cite{yat96a,dha96a}, YbRh$_2$Si$_2$ \cite{tro00a}, YbAgGe \cite{bud04a,ume04a}, and more recently, the six, isostructural Yb$T_2$Zn$_{20}$ ($T$ = Fe, Co, Ru, Rh, Os, Ir) \cite{tor07a} compounds.

Fragile magnetism in Yb-based compounds can be searched for, and studied, in several ways.  When systems order, the ordering can be perturbed, and, in some cases, driven toward a $T = 0$ K, quantum phase transition.  When the phase transition can be maintained as a second order one this will result in a quantum critical point (QCP).  When the unperturbed system does not manifest magnetic order upon cooling, i.e. when it crosses fully over to a heavy Fermi-liquid state, then a phase transition can sometimes be induced by pressure.  In either case, chemical substitution can be used to perturb the system.  Substitution is often a very powerful perturbation, though, and introduces complications due to disorder that can then make analysis complicated, e.g.: increased scattering, distribution of Kondo-temperatures and / or CEF splittings, loss of coherence, just to name a few. Keeping this in mind, the use of pressure or magnetic field on highly ordered compounds is preferred.

Figures 1a and 1b present the $H-T$ phase diagrams for YbAgGe and YbBiPt respectively.  Both of these compounds have reduced moment ordering occurring after the sample is cooled below $T_K$. In both cases the $H = 0$ T ordering occurs at relatively accessible temperatures and the magnetic field scales are neither unreasonably small, nor large, allowing for measurement in standard, lab-based systems.  The YbAgGe $H-T$ phase diagram has evolved over the past decade as the result of a slew of thermodynamic, transport and microscopic measurements \cite{bud04a,fak05a,bud05a,bud05b,tok06a,nik06a,fak06a,mun10a,nak11a,sch11a,tok13a}.  For the magnetic field applied in the easy plane of this hexagonal material there is a cascade of at least three ordered states, labeled a,b,c in Fig. 1a, followed by phase d, which has a linear, low temperature resistivity over the whole field range.  This is followed by phase e, over which the power law of the low temperature resistivity changes from linear to $T^2$, and finally, for fields greater than $\sim 12$ T, phase f, which manifests Fermi-liquid like behavior.  Some of the current open questions associated with YbAgGe are: what is the nature of the d-phase and how do magnetic fluctuations evolve across the potential bicritical point, QCP and toward the ultimate Fermi-liquid region.  YbAgGe combines field tuned quantum criticality with possible Kondo effect exacerbated frustration (i.e. frustration in the Yb compound that is apparently absent or less significant in the $R$AgGe analogues for $R$ = Tm - Tb) \cite{mor04a} and reveals a rich and complex $H-T$ phase diagram.  Whereas, ultimately, we may learn a lot from this system, the guaranteed entropy is taking the long way home in terms of how it gets removed.

YbBiPt on the other hand, appears to be a classic QCP system.  Although YbBiPt was discovered to be a HF system over 20 years ago \cite{can91a,fis91a}, its proximity to quantum criticality as well as the microscopic details of its presumed magnetic order were only determined over the past few years \cite{mun13a,mun15a,uel14a,uel15a}.  As shown in Fig. 1b, an applied magnetic field of roughly 0.5 T is enough to suppress the thermodynamic and transport features associated with long range magnetic order.  By roughly 1 T there is a clear, low temperature Fermi liquid (FL) state with diverging values of the coefficient of $T^2$ resistivity and linear specific heat as $H$ is decreased from higher values \cite{mun13a}.  At much higher fields the heavy Fermi-liquid state is suppressed; by 7 T clear quantum oscillations are detected with electron masses all of order roughly unity \cite{mun15a}.  Until recently, the missing, key data set, was detailed, microscopic, information about the magnetic ordering.  Significantly larger single crystals, combined with two decades worth of improvements in detectors recently allowed for the determination of the ordering wave vector $\left(\frac{1}{2}~\frac{1}{2}~\frac{1}{2} \right)$ and moment direction [1~1~1] \cite{uel14a}.  It is very important to note, though, that the measured diffraction data consists of a broader peak (with an estimated coherence length of $\sim 20$ \AA) and a sharper peak (with a coherence length of $\sim  80$ \AA).  Whereas the sharper peak appears at $T_N \sim 0.4$ K, the broader peak gradually appears as $T$ drops below 1 K.  Although it might be tempting to dismiss the relative broadness these peaks as being related to some sort of sample disorder or imperfection, this is hard to do when the same samples show exquisitely clear quantum oscillations for the relatively modest fields of 7 T and higher \cite{mun13a,mun15a}. 

Given the proximity of YbBiPt to its QCP, even at $H = 0$, it is possible that the broad peak is associated with quantum critical fluctuations that are already slowing down at low temperatures.  At this point, YbBiPt stands poised to allow direct testing of many of the predictions associated with QCPs.  Inelastic scattering studies for fields up to 5 T and for temperatures in the 50-1000 mK range will reveal how the removal of the residual entropy associated with the $4f$-shell CEF split ground state changes across the critical region.

Yb-based HF compounds that do not manifest any sign of magnetic order can still be probed for fragile magnetism by the application of pressure or strain.  At the simplest level, pressure can be viewed as pushing the samples from the Fermi-liquid part of the Doniach phase diagram toward the magnetically ordered side of the putative QCP.  Such behavior can be seen in YbFe$_2$Zn$_{20}$ \cite{kim13a}, one of the 6 structurally related Yb-based heavy fermions discovered in the Yb$T_2$Zn$_{20}$ family \cite{tor07a}. YbFe$_2$Zn$_{20}$, at ambient pressure, has an electronic specific heat, $\gamma$, of 520 mJ / mole~K$^2$, a $T_K \sim 33$ K, and from a generalized Kadowaki-Woods (KW) analysis has the full $J = 7/2$ CEF multiplet contributing to the low temperature FL state.  Figure 2  shows the temperature dependent resistivity of YbFe$_2$Zn$_{20}$ for representative pressures and figures 3a-3c show the pressure evolution of the temperature below which the resistivity is proportional to the Fermi-liquid, $T^2$ power-law as well as the prefactor of the $T^2$ resistivity, $A$.  These data strongly suggest that by 10 GPa there will be a pressure induced quantum critical point \cite{kim13a}.  The tantalizing question is what will be the nature of the ordered state on the high side of this QCP: antiferromagnet, ferromagnetic, or possibly some multipolar ordering?  It should be noted that, on one hand, tuning the ground state with pressure is generally more difficult than tuning with applied magnetic field.  On the other hand, if a QCP can be reached, then the chances of finding the elusive Yb-based, HF superconductor are much higher since there will not be an applied field suppressing the superconducting (SC) state.
\\

In each of these examples the assured entropy associated with a Kramers ion has made it possible to look for some sort of quantum phase transition, and, in these cases, very likely a quantum critical point.   Once we leave Yb- and Ce-based compounds and move to non-Kramers rare earth ions, we lose this degree of surety.  This is what happens when Pr-based compounds are studied for possible fragile or Kondo-like states.  

Pr, like Ce, can have tri- and tetravalent ionization.  Pr$_4$O$_{11}$ is a mixed oxide, with both states formally present.  In addition, if a Pr$^{3+}$ ion can be coaxed into a hybridizing, low-temperature state, then it will have the same CEF splitting as U$^{4+}$, so much of the theoretical machinery developed for U-base heavy fermions may be applicable to potential, hybridizing Pr-based systems.  All of this is by the way of motivating the examination of Pr-based systems for novel states that emerge from preserved entropy.  

The discovery of a heavy fermion state in PrAg$_2$In \cite{yat96b}was the result of a search for a Pr-based compound with a $\Gamma_3$ CEF split ground state.  The $\Gamma_3$ state is a non-magnetic doublet that occurs when the Pr ion is in a cubic point symmetry.  This, then, requires a compound with a cubic unit cell that, in addition, has the Pr-ion in a cubic point symmetry.  Even if the  $\Gamma_3$ CEF level is the ground state, entropy can be removed by a cooperative Jahn-Teller transition, as is found at $\sim 0.4$ K in PrPb$_3$ \cite{mor82a}, one of the few other known Pr-systems with a known  $\Gamma_3$ ground state.  (This is similar to the case of Ce-based compounds where there is a race between Kondo screening and magnetic order to remove entropy; in Pr-based compounds the race includes structural transitions that can change the point symmetry of the Pr-ion and thereby lead to a singlet ground state with no degeneracy.)  

PrAg$_2$In has a  $\Gamma_3$ CEF ground state that is well separated (6.1 meV) from the first excited,  $\Gamma_4$ triplet, state and has the remaining  $\Gamma_5$ triplet and  $\Gamma_1$ singlet located at 8.1 and 15.2 meV respectively \cite{gal84a,kel00a}.   No sign of any phase transition has been found for $T > 50$ mK.  Upon cooling, the magnetic specific heat can be well fit, for $T > 10$ K, to Schottky anomalies associated to this  $\Gamma_3$ -  $\Gamma_4$ -  $\Gamma_5$ -  $\Gamma_1$ splitting.  Below 10 K, the specific heat rises to a broad maximum near 0.5 K and then drops, linearly toward zero.  As shown in the inset of Fig. 4, the linear specific heat term, $\gamma$, rises and below $\sim 0.3$ K saturates near 7 J/mole-K$^2$.  The magnetic susceptibility is consistent with a non-magnetic CEF ground state and measurements of the specific heat up to 9 T show that there is little field dependence of this greatly enhanced, low temperature $\gamma$-value (after nuclear Schottky and its fiels dependence are acccounted for) \cite{mov99a}.  Specific heat measurements up to 9 T (Fig. 4) also show little field dependence other than that associated with the Pr - nuclear Schottky peak (see Fig. 2 of Ref. \cite{mov99a} and related discussion or details).

All of these data suggest that PrAg$_2$In enters into a strongly correlated state upon cooling and, for $T > 0.3$ K has one of the larger $\gamma$-values of any heavy fermion compounds, only being exceeded by YbBiPt \cite{can91a,fis91a,mun13a} and YbCo$_2$Zn$_{20}$ \cite{tor07a}.  Although the very large $\gamma$-value is noteworthy, what is far more significant is the fact that this exceptionally large $\gamma$-value is emerging from a non-magnetic,  $\Gamma_3$ CEF level.  At a gross level, the Kondo screened state in PrAg$_2$In can be thought of as the conduction electrons dynamically screening the Pr-ions' multipolar-moment in much the same way that, for a Ce-ion, there is a dynamic screening of the magnetic moment.  As such, PrAg$_2$In provided the first clear example of a Pr-based heavy fermion compound based on Kondo-screening of a non-magnetic state.  

PrAg$_2$In also illustrates the scheme of preserving entropy to low enough temperatures so as to allow exotic states to emerge.  Although Pr-hybridization could possibly allow for a more traditional Kondo screening of a moment bearing CEF level, this is energetically less likely for two reasons: (1) the hybridization in Pr-systems is much weaker than in Ce-based ones, so the scale of $T_K$ is greatly reduced and (2) the de Gennes factor for Pr is $\sim 4$ times larger than for Ce, setting a high temperature scale for RKKY mediated local moment ordering.  Both of these energy scale changes relative to Ce make the  $\Gamma_3$ CEF state the safest place to stash entropy for potential low temperature "use".  Further searches for Pr- and Tm-based heavy fermions are important, but are also difficult, given their weaker hybridization and non-Kramer’s status. For Pr-based systems PrMg$_3$ \cite{gal81a,tan06a,mor09a,ara11a} was also studied, and more recently, based on the wealth of physics found in the $R$$T_2$Zn$_{20}$ families \cite{tor07a,jia07a,jia07b,jia08a,jia09a,nin11a} with the $R$-site in cubic point symmetry, there have been studies of Pr$T_2$Zn$_{20}$, Pr$T_2$Cd$_{20}$, and Pr$T_2$Al$_{20}$ systems \cite{oni10a,oni12a,iwa13a,yaz15a,sak11a,tsu14a}.

Currently, first principles, computational studies provide very little (if actually any) guidance to the CEF splitting or CEF split ground state of the Hund's rule $J$ multiplet for intermetallic compounds.  This means that, once a compound with the right unit cell class, the right number of unique crystallographic sites for the rare earth (almost always this number is unity), and the right point symmetry for the rare earth site, is found then experimental data is needed to infer the CEF splitting and CEF ground state.  Given the relative dearth of examples of Pr- or Tm-based heavy fermion compounds, such a search is needed if we are to better understand their possible, hybridized ground states.
\\

Although the scheme outlined above for searching for Pr-based HF systems is much harder that associated with similar searches for Yb- or Ce-based systems, all of these examples, so far, are springing off of the more or less guaranteed entropy associated with each of these rare earths' Hund's rule ground state multiplet, $J$.  In terms of preserving entropy to low enough temperatures, each of these cases at least starts with a well understood starting entropy.  Now, as we shift our attention to transition metal based fragile magnetism, even this starting entropy becomes far from guaranteed.

First, we need to elaborate a little on what we mean by "fragile magnetism" in this situation.  We do not mean that the samples are brittle and can shatter when dropped (this has actually been a question at a recent colloquium).  What we mean by this term is that via some perturbation, be it applied pressure, applied magnetic field or chemical substitution, a magnetic transition, and its associated fluctuations and entropy, can be brought to low enough temperatures to allow (hopefully) for a new phase, or state, to emerge.  A canonical example of this is illustrated in Fig. 5 where the $T-x$ and $T-P$ phase diagrams for BaFe$_2$As$_2$ are shown \cite{nin08a,col09a}.  For the case of Co-substitution, Ba(Fe$_{1-x}$Co$_x$)$_2$As$_2$, for $x = 0.00$ there is an orthorhombic (Ortho) phase transition followed closely by antiferromagnetic (AF) ordering for $T \sim 135$ K.  The size of the ordered moment in BaFe$_2$As$_2$ is 0.87 $\mu_B$ \cite{hua08a}.  As $x$ is increased the AF ordering temperature is suppressed and the size of the ordered moment is reduced.  By $x = 0.047$ the AF transition occurs near 47 K and the size of the ordered moment has been reduced to 0.2 $\mu_B$ and the system becomes superconducting (SC) below a $T_c$ of 17 K \cite{pra09a}.  For $x = 0.063$ there is no longer any AF transition and the superconducting transition has its maximum value of $\sim 25$ K \cite{nan10a}; as $x$ increases further $T_c$ drops and goes to zero by $x \sim 0.15$.  The antiferromagnetic phase transition in BaFe$_2$As$_2$ is an example of what we mean to capture by the phrase, "a fragile magnetic phase transition".  Not only is the transition temperature suppressed, but he size of the ordered moment decreases as $T_N$ drops.   A similar phase diagram is found for pure BaFe$_2$As$_2$ under pressure (Fig. 5b).  As pressure is increased the AF ordering drops and a superconducting dome emerges with a maximum $T_c$ near the point where the $T_N(P)$ line extrapolates to zero \cite{col09a}.  

CaFe$_2$As$_2$ \cite{nin08b,ron08a,wug08a,can09a} provides an even more extreme example of "fragility".  Figures 6a and 6b present $T-x$ and $T-P$ phase diagrams (respectively) for CaFe$_2$As$_2$ \cite{ran12a,gol09a}.  As Co is substituted for Fe the antiferromagnetic / orthorhombic phase transition is suppressed to 0 K and for higher substitution levels the "over-doped" half of a superconducting dome is revealed.  Unlike BaFe$_2$As$_2$, CaFe$_2$As$_2$ has a single, first order, coupled magnetic and structural phase transition that, over the substitution and pressure ranges shown, does not split into separated second order phase transitions \cite{ran12a,gol09a,gol08a,yuw09a}.   It should be noted, though, that although this first order phase transition is associated with a very sharp, almost full jump of the order parameters to their low temperature saturated values at the transition temperature \cite{gol09a,gol08a,kre08a}, there are very clear and long-lasting (to higher temperatures) magnetic fluctuations above the phase transition \cite{pra09b,dia10a,soh13a}.  The sudden loss of fluctuations as the sample jumps deep into the AF/Ortho state (via the first order phase transition) is thought to be the reason for the absence of SC on the "underdoped" side of the phase diagram.  

The pressure dependence of pure CaFe$_2$As$_2$ (Fig. 6b) reveals just how fragile the magnetism in CaFe$_2$As$_2$ is.  The 170 K, ambient pressure phase transition is suppressed to 100 K by 0.4 GPa and for higher pressures a first order phase transition to a non-magnetic, collapsed tetragonal phase rapidly rises, and reaches 300 K by 1.6 GPa.  The discovery \cite{kre08a,tor08a} and understanding \cite{gol09a,dha14a,fur14a,die14a} of the cT phase transition illustrates the Janus-like nature of very fragile magnetism:  looking at one side, this means that the system can be tuned and controlled readily in a number of ways; looking at the other side, this means that the magnetism can simply disappear if care and diligence are not taken.  Indeed, CaFe$_2$As$_2$ is so very pressure and strain sensitive that post-growth annealing and quenching of single crystals grown out of FeAs solutions can lock in enough strain to stabilize the tetragonal to collapsed tetragonal phase transition at ambient pressure, completely bypassing any transition to a magnetically ordered state \cite{ran11a}.

A three-dimensional phase diagram can be made for Ca(Fe$_{1-x}$Co$_x$)$_2$As$_2$ at ambient pressure (Fig. 7a) that illustrates the interplay between the four salient ground states: AF/ortho, SC/tet, paramag/tet, non-magnetic (quenched moment) /cT \cite{ran12a}.  To further illustrate the equivalence of the effects of post growth annealing and quenching and physically applied, hydrostatic pressure, Fig. 7b shows an $x = 0.028$ sample that was annealed/quenched at 350$^\circ$ C and has and ambient pressure AF/ortho phase transition at 50 K \cite{gat12a}.  Very modest pressure of 30 MPa (0.03 GPa or 0.3 kbar) suppresses the AF/ortho phase transition to zero and reveals a SC phase with $T_c \sim 16$ K.  By 170 MPa (0.17 GPa ar 1.7 kbar) the SC phase with its reduced $T_c$-value of $\sim 10$ K is suddenly replaced by the cT phase.  The effects of pressure on this $x = 0.028$ sample are qualitatively the same as can be seen in Fig. 7a for increased post growth annealing and quenching temperatures.  The pressure scale in Fig. 7b is in MPa; the pressures needed to change the low temperature state from AF/ortho to paramagnetic-SC/tet to non-magnetic/cT are less than 2 kbar (0.2 GPa).  These pressure values are comparable to those generated under the stiletto heal worn by an average sized physicist.  Perhaps more importantly, the strain caused by differential thermal contraction between the sample and a mounting surface can, and has, caused significant changes in transition temperature and even state.  The moral here being that if a compound is highly tunable, then it is also very sensitive to small, perhaps unintentional perturbations associated with growth or measurement; caveat emptor!

The fragile magnetism manifest by CaFe$_2$As$_2$ is not only remarkable, even bordering on pathological, but it also serves as another step in the progression away from the guaranteed entropy and magnetism of a Kramers-ion rare earth, to the less guaranteed entropy / magnetism of the non-Kramer’s rare earth, to far from guaranteed, or even likely, magnetism in transition metal based intermetallic systems.  In the same way PrAg$_2$In, and the related PrPb$_3$ \cite{mor82a}, can lose the entropy associated with the $\Gamma_3$ doublet via a structural phase transition that changes the point symmetry of the Pr-ion, CaFe$_2$As$_2$ can lose its Fe-based magnetism through the T-cT structural phase transition that leads to a reduction both the correlations (Coulomb $U$) and the density of states at the Fermi level, $D(E_F)$.  Both of these reductions work together to lower the Stoner parameter, $U  D(E_F)$, an lead to a loss of local moment like behavior on the Fe site \cite{die14a}.
\\

This general form of the Fe-base superconductor phase diagram is far from unique; similar phase diagrams can be found to the CuO-based high $T_c$ compounds as well.  Indeed it is the promise of higher temperature (energy) scales, as well as the possibility of finding still further examples of high temperature superconductivity that makes the search transition metal based fragile moment systems so compelling.  Unfortunately, in some abstract karmic balance, the difficulty of finding transition metal based systems that are both metallic and manifest some degree of fragile magnetism seems to increase proportionally to the potential for exciting physics or higher $T_c$-values.  Indeed, some have argued that Nature (the abstract, anthropomorphic manifestation, NOT the journal) has a death-wish for experimental physicists and solid state chemists since many of the promising and exciting systems in the post-high-$T_c$ world have increasingly involved the use of elements and compounds most commonly used as rat-poison, plot devices in murder mysteries, and bombs / fireworks.  We will come to the difficulty of this search in greater detail soon, but first, let's better delineate the phrase "fragile magnetism" some more.
\\

Although driving a magnetic transition temperature to zero is a highly desired aspect of a fragile magnet, it is not a defining property; mathematically speaking it is a necessary, but not sufficient, condition.  This is most clearly illustrated by returning, briefly, to the $4f$-shell physics of the rare earths.  Gd has a very robust local moment with no significant CEF splitting due to its Hund's rule groundstate being $J = S= 7/2$ (i.e. $ L = 0$).  This essentially means that Gd ions always have  large, isotropic, local moments.  A Gd-based intermetallic compound that orders magnetically can, generally have a  non-magnetic rare earth such as Y, Lu or even La substituted for Gd on the rare earth site.  As this happens the magnetic ordering temperature will be suppressed and ultimately head toward zero temperature, generally passing through a spin glass phase as the relative Gd content approaches zero \cite{myd93a}.  It should be noted, though, that such a suppression of the magnetic ordering temperature is not what we mean by "fragile magnetism".  If this were the case, virtually every rare earth based compound would qualify, since in most cases non-magnetic dilution is possible.  In this example of Gd - (Y/Lu/La) dilution it is anticipated that the size of the effective moment as well as the size of the ordered moment (per Gd ion) remains unchanged.  This relative invariance of the effective as well as ordered moment makes the magnetism robust rather than fragile.  Another way of thinking about this is to appreciate that, as the magnetic transition temperature is reduced, no new mode of conduction electron / magnetic moment interaction or crossover is introduced.  The local moment interaction between the conduction electrons and the Gd remains fundamentally the same.  This is to be contrasted with the Kondo systems such as the Ce- or Yb-based compounds discussed earlier.  In the case of these hybridizing compounds, if a magnetic ordering temperature can be dropped enough, the hybridization between the $4f$-shell and the conduction electrons leads to a reduced moment and an enhanced electronic specific heat component.  To some extent, the concept of "fragile magnetism" is associated with the ordering of moments that are themselves "fragile" in some manner; moments that can diminish in size and/or increasingly interact with conduction band through channels that can give rise to effects other than those simply associated with the RKKY interaction.
\\

We can see this distinction between fragile and non-fragile, transition metal magnetism in what, at cursory glance, might appear to be rather similar systems:  LaCrSb$_3$ and LaCrGe$_3$ \cite{har97a,raj98a,bie07a}.  Both of these compounds becomes ferromagnetic near 100 K, with $T_C = 132$ K for LaCrSb$_3$ and $T_C = 85$ K for LaCrGe$_3$.  The vanadium analogues to each compound:  LaVSb$_3$ and LaVGe$_3$, are non-moment bearing, Pauli paramagnets.   Figures 8a and 9a show the pressure dependence of $T_C$ for LaCrSb$_3$ and LaCrGe$_3$ respectively \cite{lin14a,tau16a,kal16a}.  Whereas the Curie temperature of LaCrSb$_3$ is essentially invariant for pressures up to nearly 6 GPa, for LaCrGe$_3$ $T_C$ drops precipitously, going to zero near 2 GPa with a probably antiferromagnetic phase transition appearing above $T_C$ near 50 K and 1.6 GPa pressure and itself going to zero near 6 GPa.  

The effects of substituting non-magnetic V for Cr can also be studied for signs of fragile, or non-fragile magnetism.  For La(Cr$_{1-x}$V$_x$)Sb$_3$, as $x$ increases the clear signature of ferromagnetic ordering changes to one more compatible with a complex, or cluster glass-like ordering which still manifests (i) hysteresis and (ii) large saturated or saturating values in low temperature $M(H)$ plots \cite{lin14a}.  Figures 8b,8c present the evolution of (i) the paramagnetic, Curie-Weiss temperature, $\Theta$, (ii) the paramagnetic effective moment, and (iii) the saturated (or lower limit for saturating) moment in the low temperature state as a function of vanadium content, $x$.  For all intents and purposes, the evolution of all three of these quantities for La(Cr$_{1-x}$V$_x$)Sb$_3$ as $x$ increases is very similar to what would be expected of their evolution in a (Gd$_{1_x}$Lu$_{x}$)$XY$ ($X$ and $Y$ being non-moment bearing second and third elements) intermetallic compound.  In other words, Cr is acting like a robust, definitely non-fragile moment.  Both the pressure dependence and the substitution dependence of LaCrSb$_3$ rank it as a transition metal based, intermetallic compound with non-fragile magnetism.  

For La(Cr$_{1-x}$V$_x$)Ge$_3$, single crystalline samples with $x < 0.22$ and pure LaVGe$_3$ ($x = 1.00$) could be grown and studied \cite{lin13a}.  As $x$ increased from $x = 0.00$ for pure LaCrGe$_3$ toward $x = 0.21$ not only does the Curie temperature drop by over a factor of two, but, as shown in Fig. 9b, (i) the paramagnetic, Curie-Weiss temperature, $\Theta$, (ii) the paramagnetic effective moment, and (iii) the saturated moment in the low temperature, ordered state all decrease significantly.  Both $\Theta$ and the saturated moment drop so dramatically that they appear to extrapolate to zero in the $0.2 < x < 0.3$ region.  Taken together with the relatively large pressure dependence of the low temperature magnetic state, LaCrGe$_3$ appears to be an example of a transition metal based compound that does manifest a fragile state.  Indeed, ongoing measurements up to pressures of 7 GPa and applied magnetic fields as high as 14 T indicate that LaCrGe$_3$ has a “wing-structure” in the $T-H-P$ phase diagram \cite{kal16a} similar to what was first found for ferromagnetic UGe$_2$  \cite{tau10a}.

LaCrSb$_3$ and LaCrGe$_3$ provide a clearly contrasting pair of Cr-based ferromagnets with very similar stoichiometries, transition temperatures, and even saturated moments (although LaCrSb$_3$ does have a significantly larger effective moment than LaCrGe$_3$).  Although experimental measurements of either pressure and/or substitution effects reveal fragility in the magnetic state found in LaCrGe$_3$ and robustness in the magnetic state found in LaCrSb$_3$, such measurements take months of experimental effort as well as elements, crucibles, furnace time, pressure cells and cryogens.  
\\

Clearly it would be a great step forward if focused computational studies could reliably indicate some degree of fragility.  The key question then is, what criterion can be both useful and readily calculated?  Figures 5 - 9 provide a potential answer:  fragile magnets appear to be fragile both in terms of substitution and pressure, at the grossest level, to band filling and band width, respectively.  This means that instead of having to contend with the complexity or approximations associated with substitution, fragile magnetism can be tested by the application of either positive or negative pressure {\it in silica}.  The determination of the ordered moment as a function of pressure may be the most effective method, although if determinations of ordering temperature, or size of paramagnetic, fluctuating moment as a function of pressure are possible, then these would be fine too.  In theory, this is a task that should be computationally tractable with only crystal structure as the input.  Clearly any further information about the nature of the ambient pressure magnetic ordering will only facilitate and refine the computational work.  The $A$Fe$_2$As$_2$ ($A$ = Ca, Sr, Ba) series as well as the LaCr$X_3$ ($X$ = Sb and Ge) set of compounds offer clear bench marks for testing and refining of methodologies.  Materials such as ZrZn$_2$ \cite{way69a,smi71a} and Sc$_3$In  \cite{gar68a,gre89a} can also serve as tests.   In addition we are currently working on other transition metal based systems, getting pressure dependencies of magnetic ordering so as to widen the set of compounds that can be used to refine, test and calibrate potential computational methodologies.

Before declaring victory, though, we need to appreciate that finding metallic, transition metal based fragile magnetic systems consists of two parts, (1) finding metallic, transition metal based systems that are magnetic (or close to being magnetic) and (2) determining which of them may be fragile.  We have discussed the second part in some detail and even have outlined how computational methods could have an impact on this search, but this still leaves the first, key step to discuss and appreciate.  Coming from a background of rare earth magnetism with occasional visits to the realms of permanent magnet and oxide physics, it was a bit of a shock just how uncommon transition metal based, compounds that are metallic and magnetic (as in moment bearing) actually are.  (At some level this is a blind spot many of researchers in this field share as can be witnessed by the apparent assumption that Ni was probably moment bearing in the RNi$_2$B$_2$C compounds when they were discovered \cite{iku94a,koh95a}.)  Once you start paying attention, though, it becomes more and more apparent that even $3d$-shell transition metals (with the conspicuous exception of Mn) are reluctant to be moment bearing except in compounds with very high relative concentrations of Cr, Fe, Co and Ni.  To make this point clear, for the $RT_2$Ge$_2$ and $RT_2$Si$_2$ series ($T$ = Fe, Co, and Ni) iron, cobalt and nickel are non-moment bearing (with a possible exception of iron in LuFe$_2$Ge$_2$ \cite{avi04a,fer06a,fuj07a,ran11b}).

In order to find transition metal based, metallic systems that manifest fragile magnetism, a superset of transition metal based metallic systems that manifest magnetic behavior is a prerequisite.  This may be the rate limiting step.  We can take ternary compounds containing Cr, Fe, Co and/or Ni as an example.  There are roughly 11,000 entries in Pearson's Crystal Data \cite{pea15a} when the search is limited to ternary compounds excluding ones with oxygen (often insulators), Be or Tl (very toxic), and $4f$ (very non-fragile, local moment magnetism) and $5f$ (radioactive) elements.  Based on experience with subsets of this list, out of these $10^4$ entries, somewhat over half of this number can be discarded as multiple entries, alloys, disordered systems, etc., leaving somewhere around 2,000 - 4,000 distinct, stoichiometric compounds to sort through.  Of these, the majority are known only as crystal structures with either no, or virtually no, temperature dependent thermodynamic or transport data that would allow for the inference of possible magnetic phase transitions.  The challenge now becomes finding which ones of these may be magnetic.  This is a Herculean task, one that again can benefit from any suggestions or guidance that computational work can offer.  

At this point we are pursuing a hybrid algorithm for deciding which of these compounds to test.  Clearly any compounds with reported phase transitions that may be magnetic rise to the top of our list.  After this, compounds that either have compelling structures or have some indication from computation that they are more stable as moment bearing, ferromagnetic systems than as non-moment bearing.   Even though this calculation is done for the $k = 0$ wave vector (ferromagnet), it often gives some indication of a predisposition to ordering, even if it is antiferromagnetic.  Hopefully over the next few years we will find a superset of metallic, transition metal based magnetic compounds that can then be reduced to a subset of fragile magnetic systems.
\\  

To wrap up, in this paper we have tried to outline some of the current ideas and themes in our own research efforts to better understand the evolution of correlated electron materials out of what we think of as fragile magnetic / preserved entropy states.  In $4f$-shell systems there are a myriad of possible exotic symmetries and degeneracies associated with the CEF splitting of the Hund's rule ground state multiplet.  Yb-based systems offer a clear test of our simplest understanding of quantum criticality, with Kramers based moments being tuned or perturbed by field, pressure, and/or doping.  Pr-based systems offer alternate mechanisms for preserving the low temperature entropy and provide examples of how CEF splitting can be used to provide novel entropy / symmetry constraints.  For both Yb- and Pr-based systems, the characteristic temperature scales will be low.  Higher temperature scales are promised by the metallic, transition metal based, fragile antiferromagents.  This class of materials is less well defined than Yb- or Pr-based compounds.  As outlined above, they are harder to find, but they are definitely out there.  The growth of their ranks and the mapping of their diversities will hopefully provide us with a clearer understanding of what are the necessary features for the development of high temperature superconductivity as well as what is the scope of other states possible as the magnetism is suppressed.  The next decade will be one of growth and exploration.  It will be an exciting and exhilarating time.  As the Hash-House Harriers would say \cite{can15a}, "On-On!" 

\ack

This work is supported by the U.S. Department of Energy, Office of Basic Energy Science, Division of Materials Sciences and Engineering and the Gordon and Betty Moore Foundation. The research was performed at the Ames Laboratory. Ames Laboratory is operated for the U.S. Department of Energy by Iowa State University under Contract No. DE-AC02-07CH11358.  Support from  the Gordon and Betty Moore Foundation came via the  EPiQS Initiative through Grant GBMF4411. PCC is grateful for long discussions with R. Valenti, M. Tomic, P. Hirschfeld, K.-M. Ho, C.-Z. Wang, R. Flint and C. Wolverton concerning computational possibilities.  PCC was able to complete work on this overview during travels supported by The Alexander von Humboldt Foundation.

\clearpage

\begin{figure}[htbp]
\begin{center}
\includegraphics[angle=0,width=100mm]{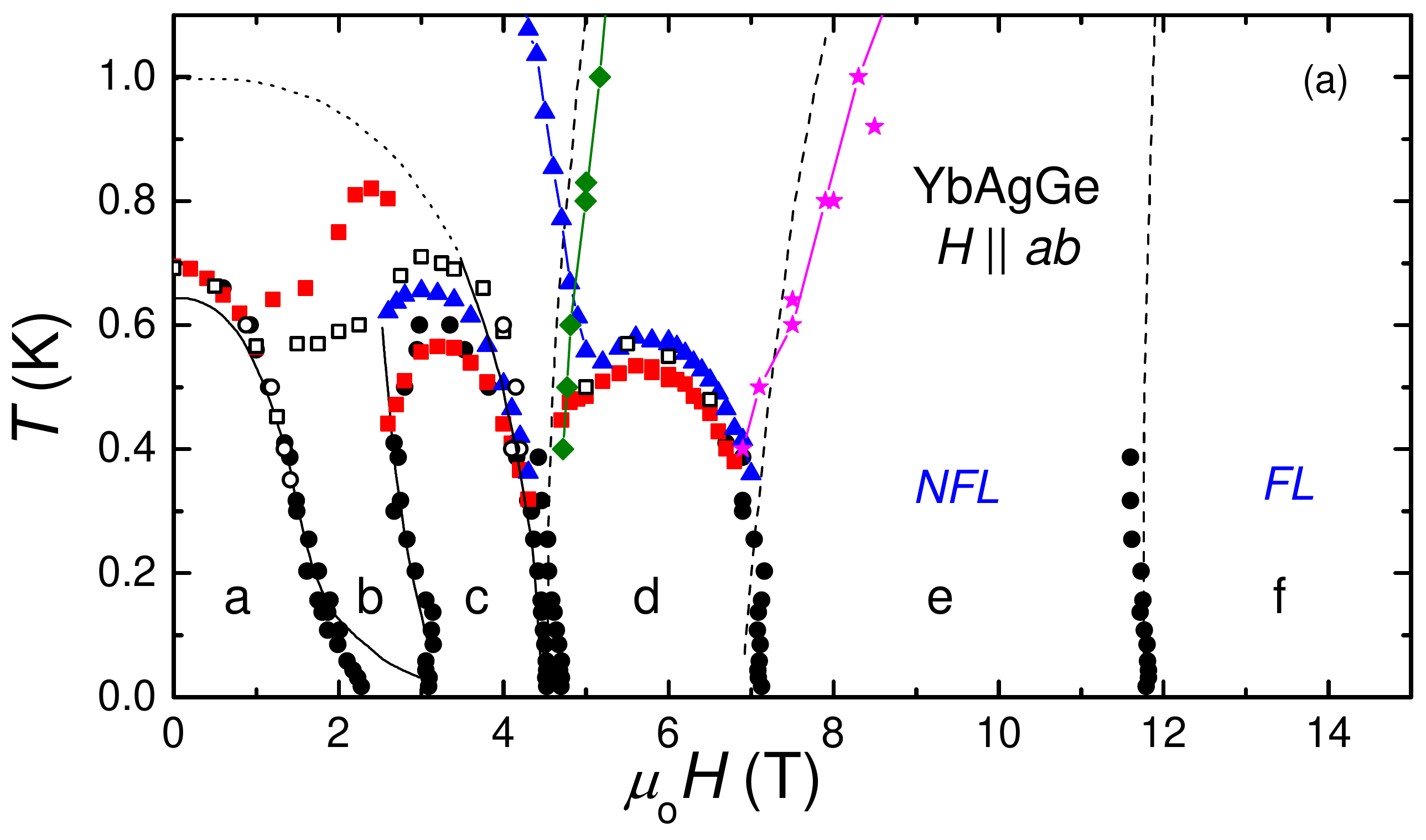}
\includegraphics[angle=0,width=110mm]{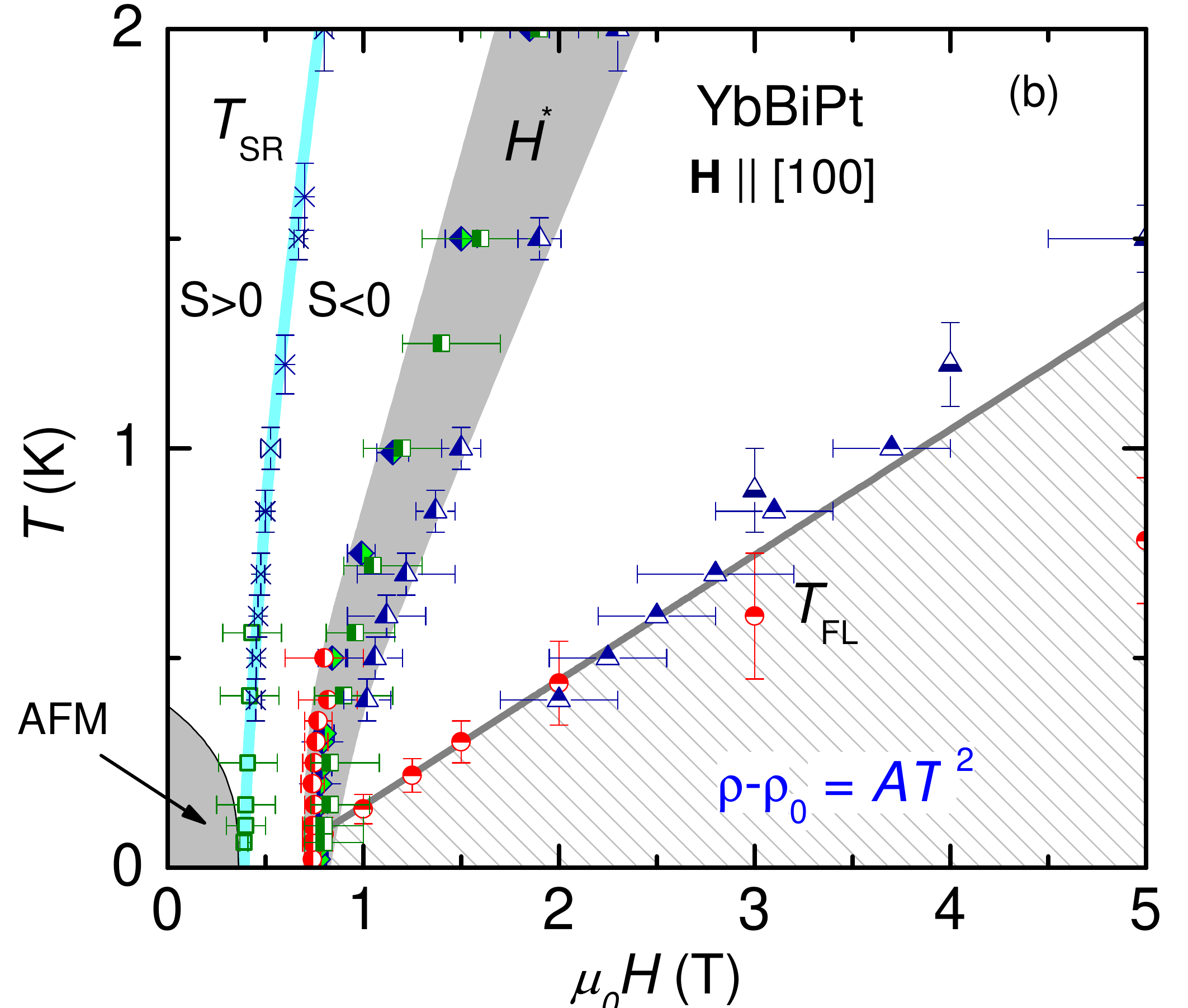}
\end{center}
\caption{(Color online) (a) The $H - T$ phase diagram of YbAgGe below 1 K with $H$ applied perpendicular to the $c$ axis; $NFL$ and $FL$ are non-Fermi-liquid and Fermi-liquid regions respectively (after Ref. \cite{sch11a}); (b) the $H - T$ phase diagram for YbBiPt; $T_{SR}$, $H^*$, and $T_{FL}$ are crossover lines corresponding to change of sign of the Seebeck coefficient, feature in Hall resistivity and the upper boundary of the $\Delta \rho \propto T^2$ behavior of resistivity respectively (after Ref. \cite{mun13a}). } \label{F1}
\end{figure}

\clearpage

\begin{figure}[htbp]
\begin{center}
\includegraphics[angle=0,width=120mm]{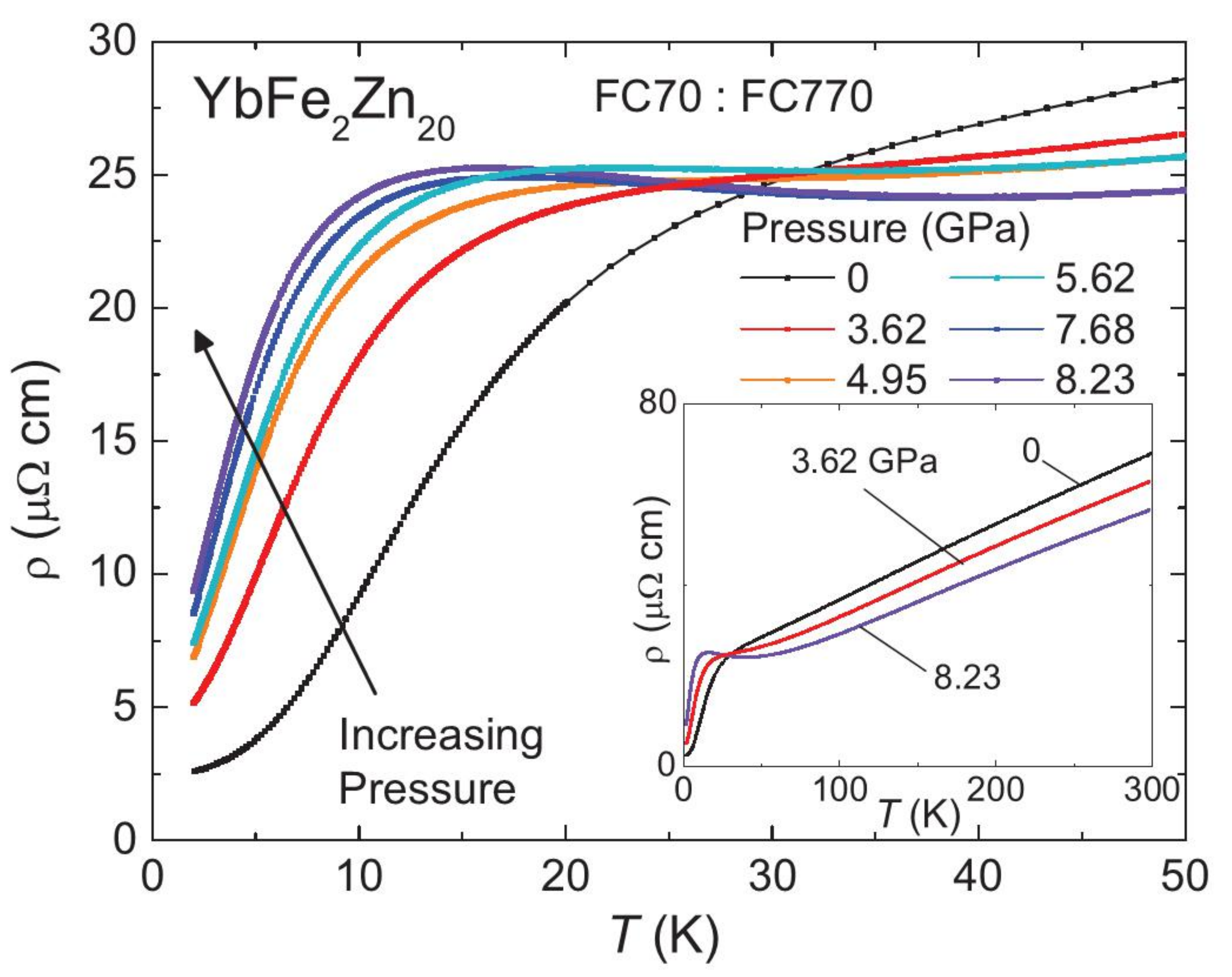}
\end{center}
\caption{(Color online) Low temperature resistivity data for YbFe$_2$Zn$_{20}$ under pressures up to 8.23 GPa. Inset: Resistivity curves up to 300 K for $P$ = 0,
3.62, and 8.23 GPa \cite{kim13a}.} \label{F2}
\end{figure}

\clearpage

\begin{figure}[htbp]
\begin{center}
\includegraphics[angle=0,width=100mm]{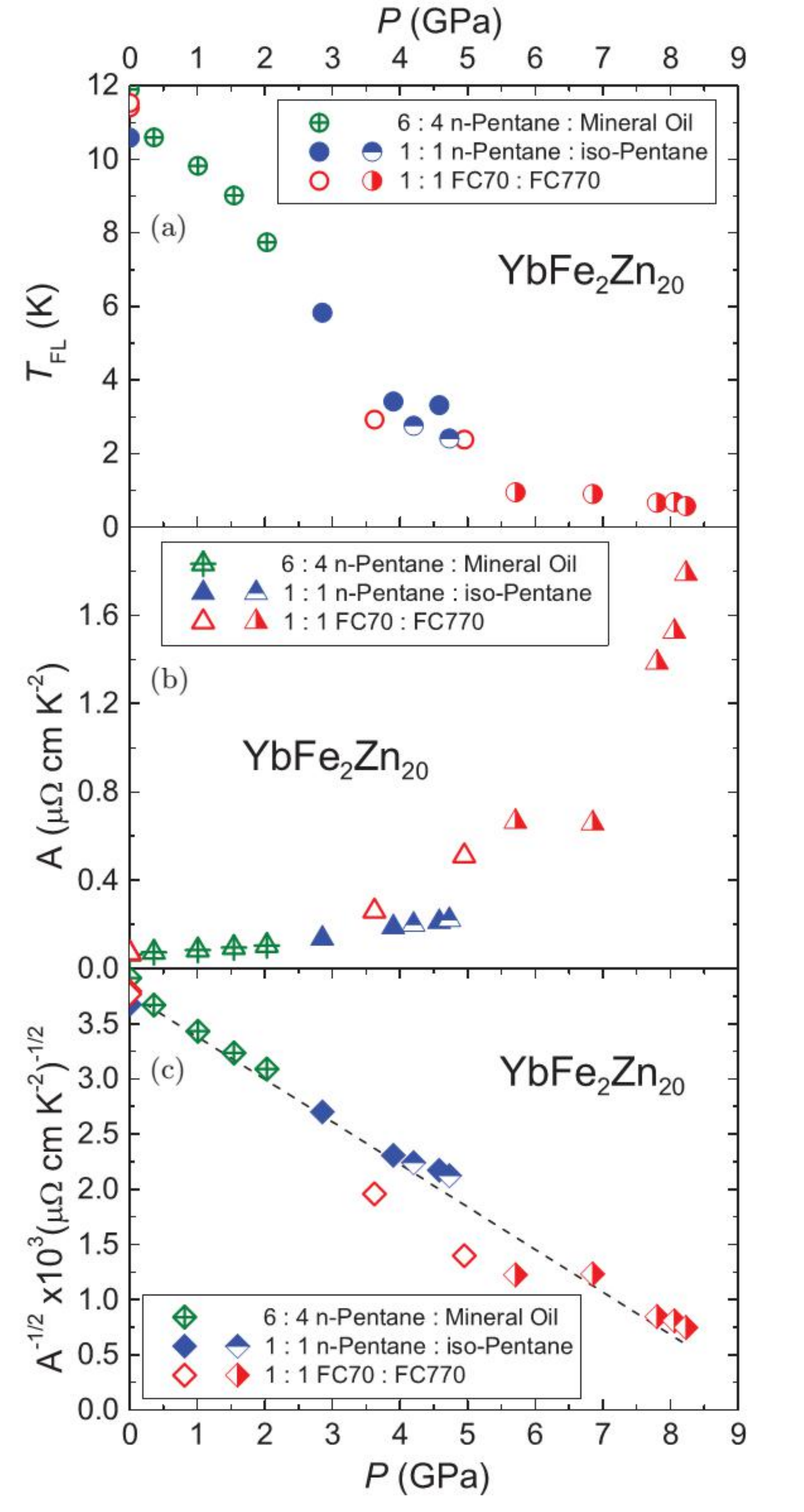}
\end{center}
\caption{(Color online) (a) Evolution of Fermi liquid temperature, $T_{FL}$, as pressure is applied; (b) divergence of the $T^2$ coefficient of resistivity under pressure; (c) $A^{-1/2}$ under pressure, the dashed line is a linear fit of the data  \cite{kim13a}. } \label{F3}
\end{figure}

\clearpage

\begin{figure}[htbp]
\begin{center}
\includegraphics[angle=0,width=120mm]{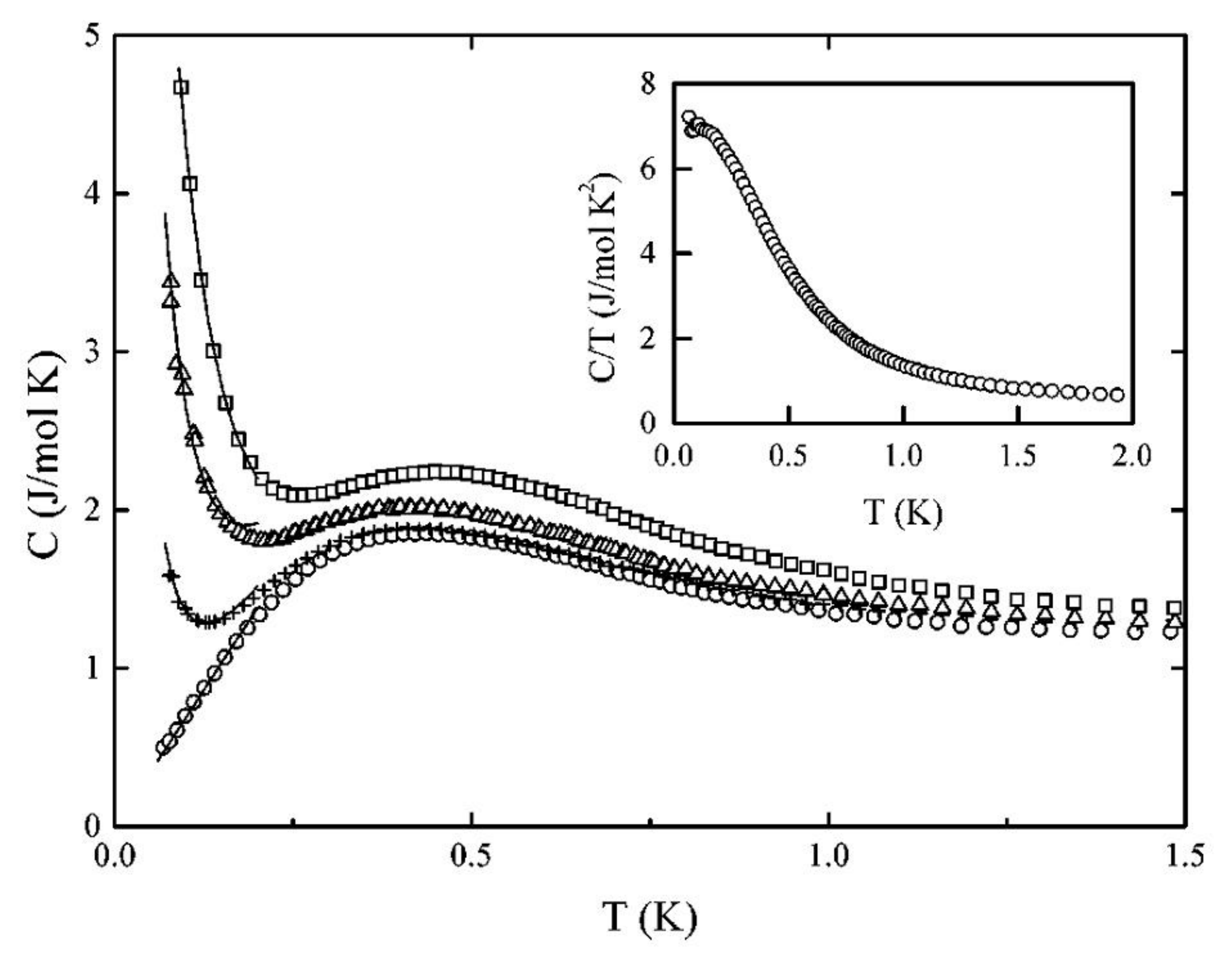}
\end{center}
\caption{Specific heat of PrInAg$_2$ in magnetic field of 0~T ($\circ$), 3~T ($+$), 6~T ($\triangle$), and 9~T ($\square$). Inset: specific heat divided by temperature for $H = 0$. Note: conspicuous, low temperature field dependence is from Pr-ion's nuclear Schottky terms \cite{mov99a}.}\label{F4}
\end{figure}

\clearpage

\begin{figure}[htbp]
\begin{center}
\includegraphics[angle=0,width=100mm]{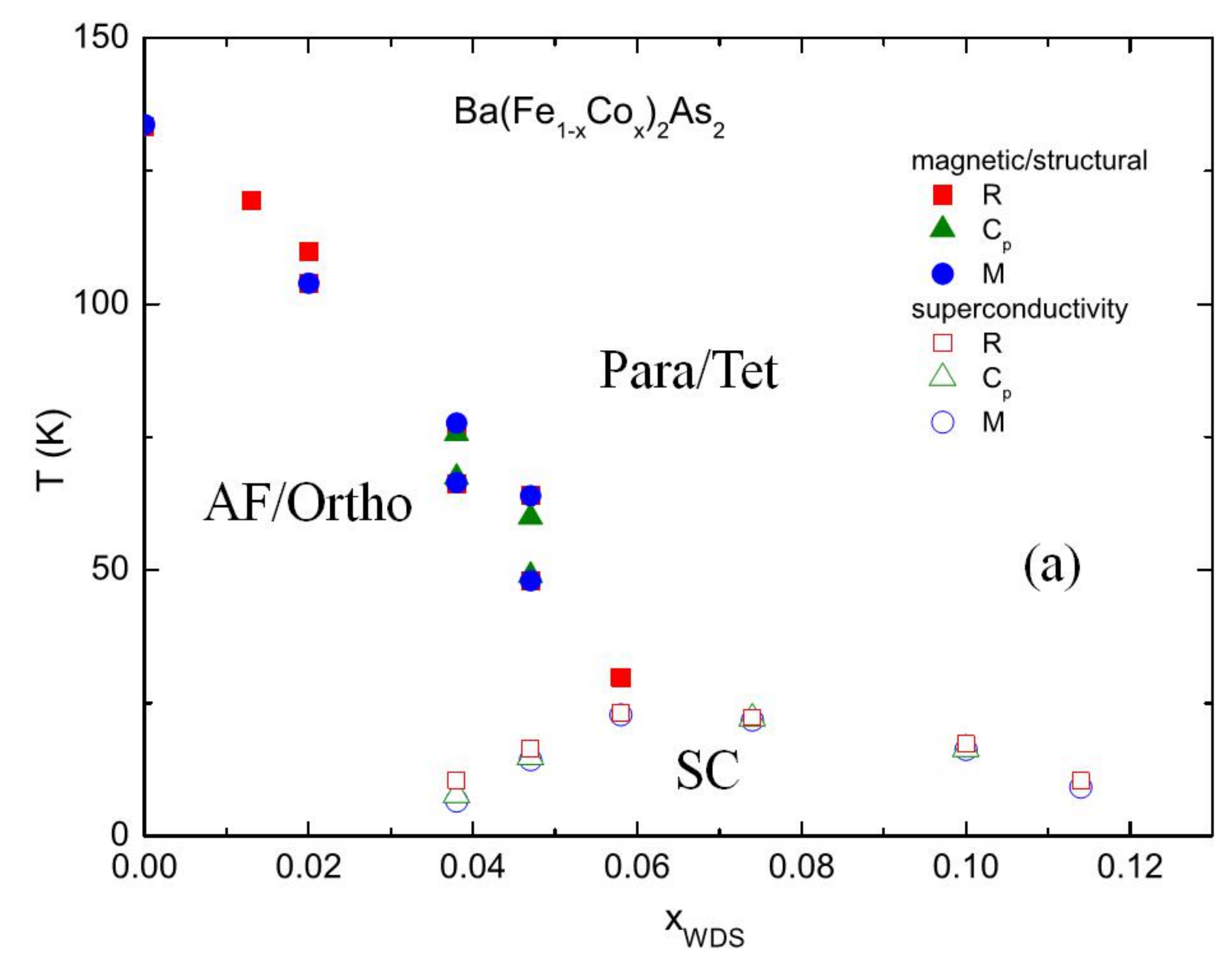}
\includegraphics[angle=0,width=100mm]{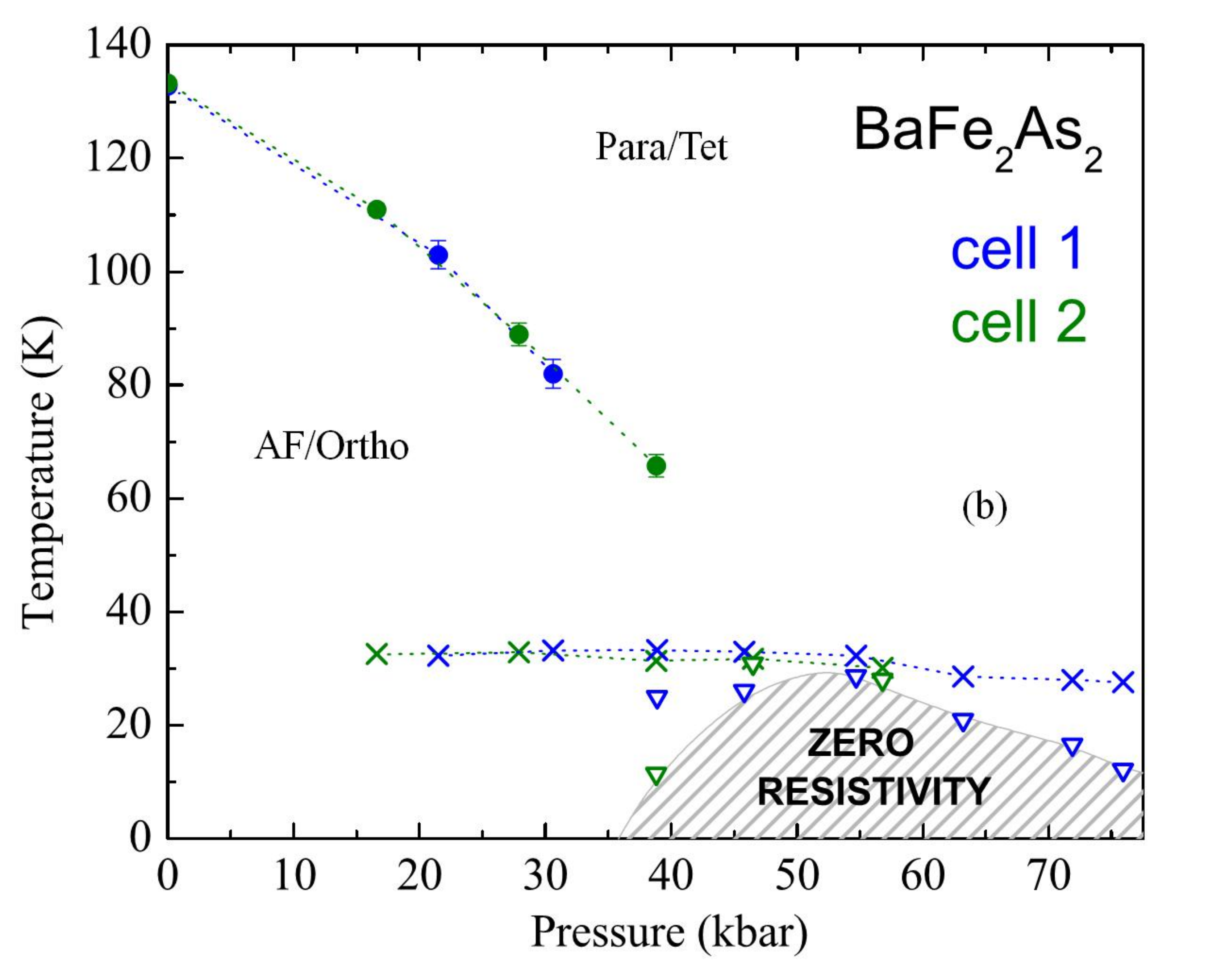}
\end{center}
\caption{(Color online) (a) $T - x$ phase diagram for Ba(Fe$_{1-x}$Co$_x$)$_2$As$_2$ single crystals \cite{nin08a}; (b)  $T - P$ phase diagram for BaFe$_2$As$_2$, the hatched area corresponds to true zero-resistance superconducting region \cite{col09a}. Paramagnetic / tetragonal, antiferromagnetic / orthorhombic, and superconducting (or zero resistivity) regions are labeled, but see Refs. \cite{nin08a,col09a} or \cite{can10a} for details. } \label{F5}
\end{figure}

\clearpage

\begin{figure}[htbp]
\begin{center}
\includegraphics[angle=0,width=100mm]{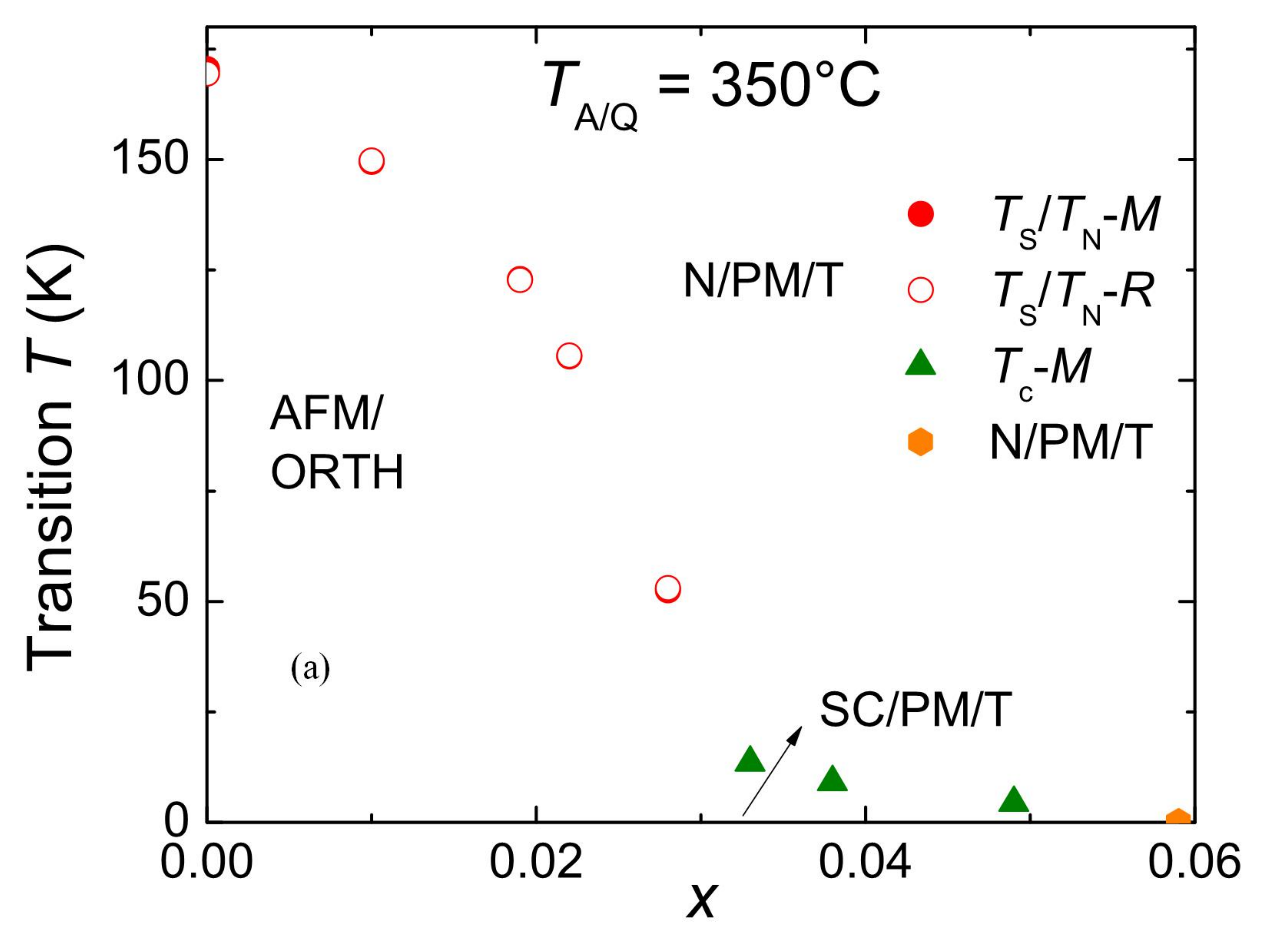}
\includegraphics[angle=0,width=100mm]{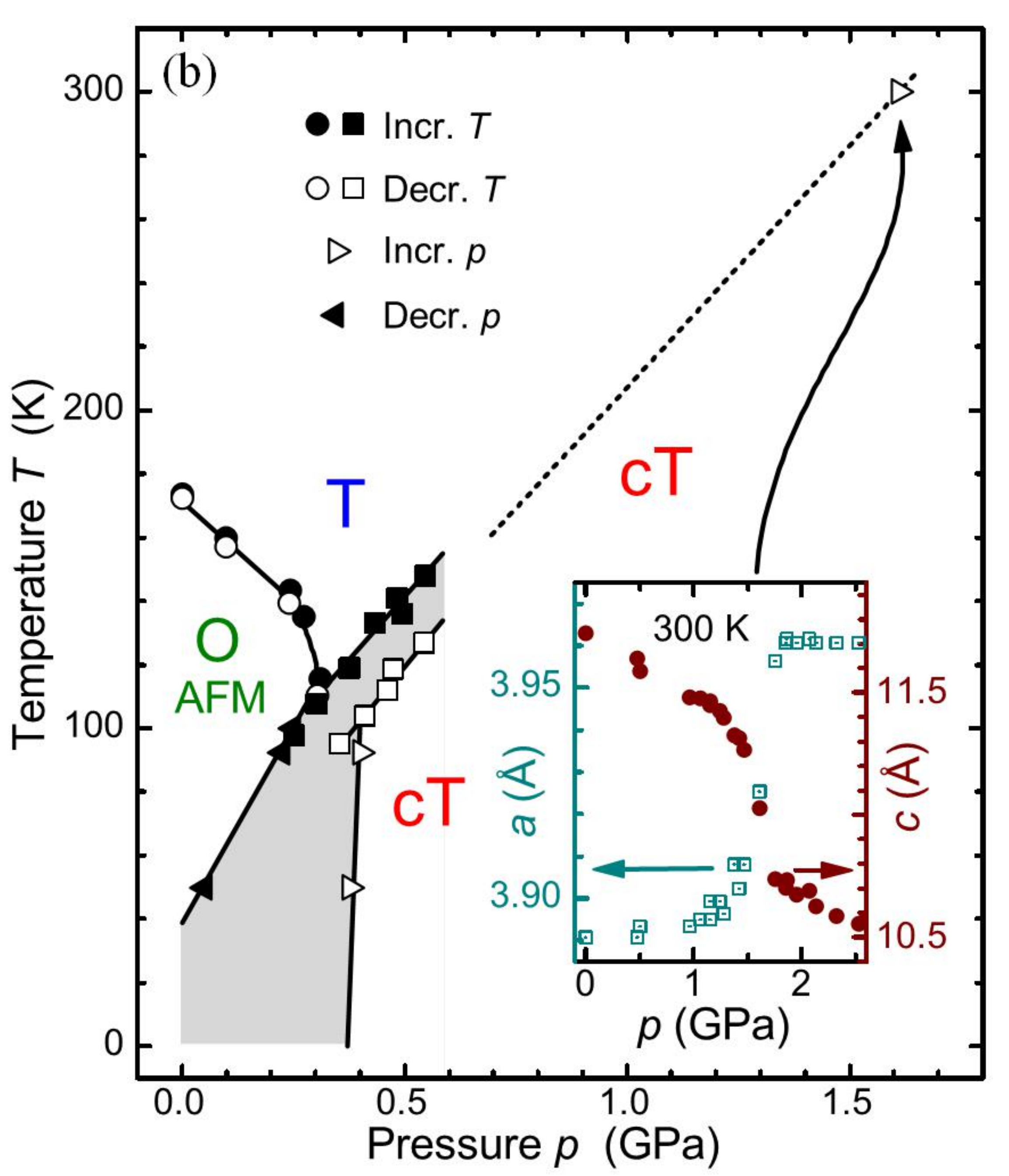}
\end{center}
\caption{(Color online) (a) $T - x$ phase diagram for Ca(Fe$_{1-x}$Co$_x$)$_2$As$_2$ single crystals with the annealing/quenching temperature of 350$^\circ$ C.  Three different phases are observed: antiferromagnetic/orthorhombic (AFM/ORTH), superconducting/paramagnetic/tetragonal (SC/PM/T), and non-superconducting/paramagnetic/tetragonal (N/PM/T) state (after Refs. \cite{ran12a,ran14a}); (b)  $T - P$ phase diagram for CaFe$_2$As$_2$, under hydrostatic pressure determined from neutron and high-energy x-ray diffraction measurements \cite{gol09a}. Tetragonal (T), collapsed tetragonal (cT), and orthorhombic / antiferromagnetic (O - AFM) phases are labeled - see Ref. \cite{gol09a} for details.} \label{F6}
\end{figure}

\clearpage

\begin{figure}[htbp]
\begin{center}
\includegraphics[angle=0,width=120mm]{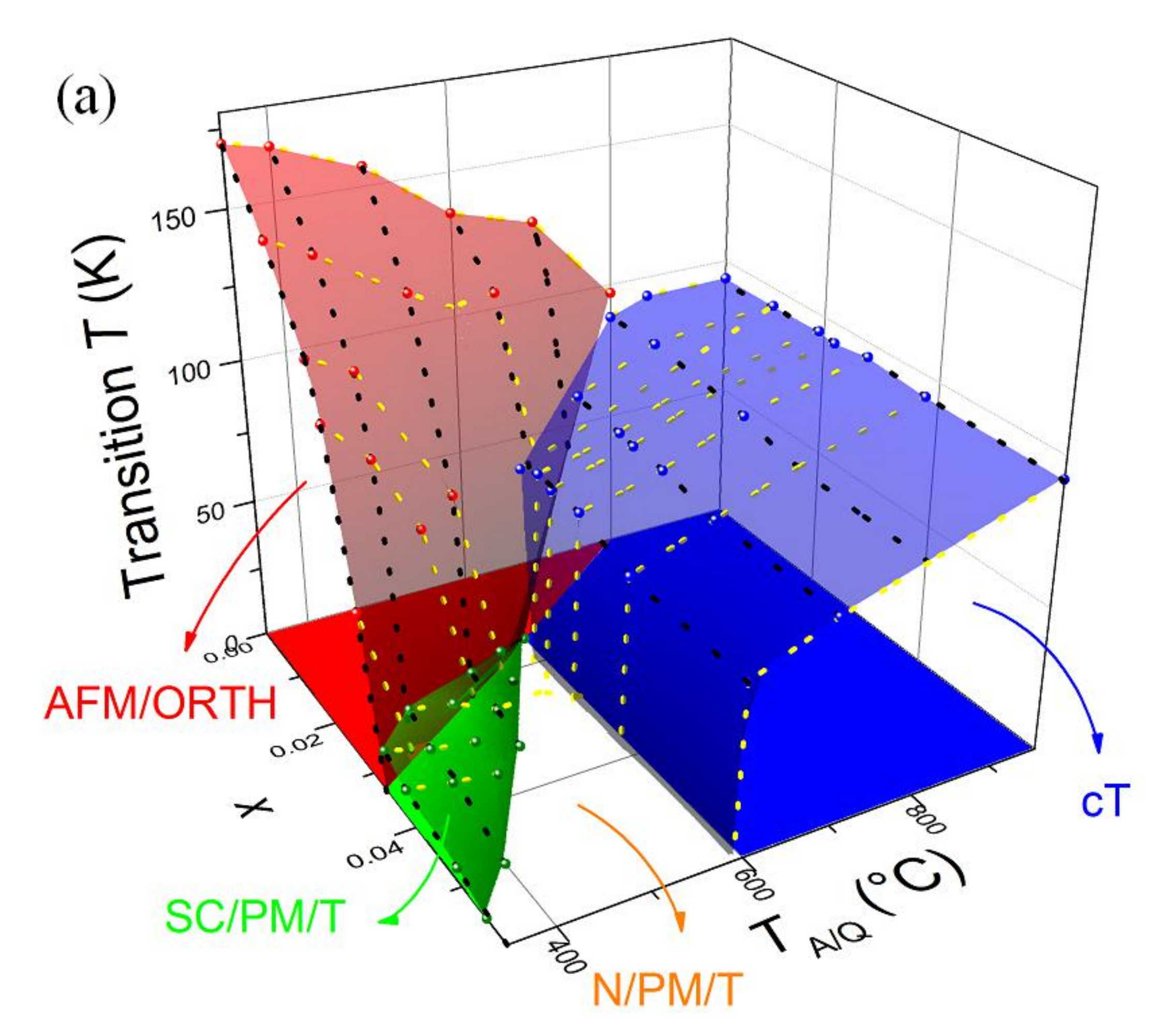}
\includegraphics[angle=0,width=120mm]{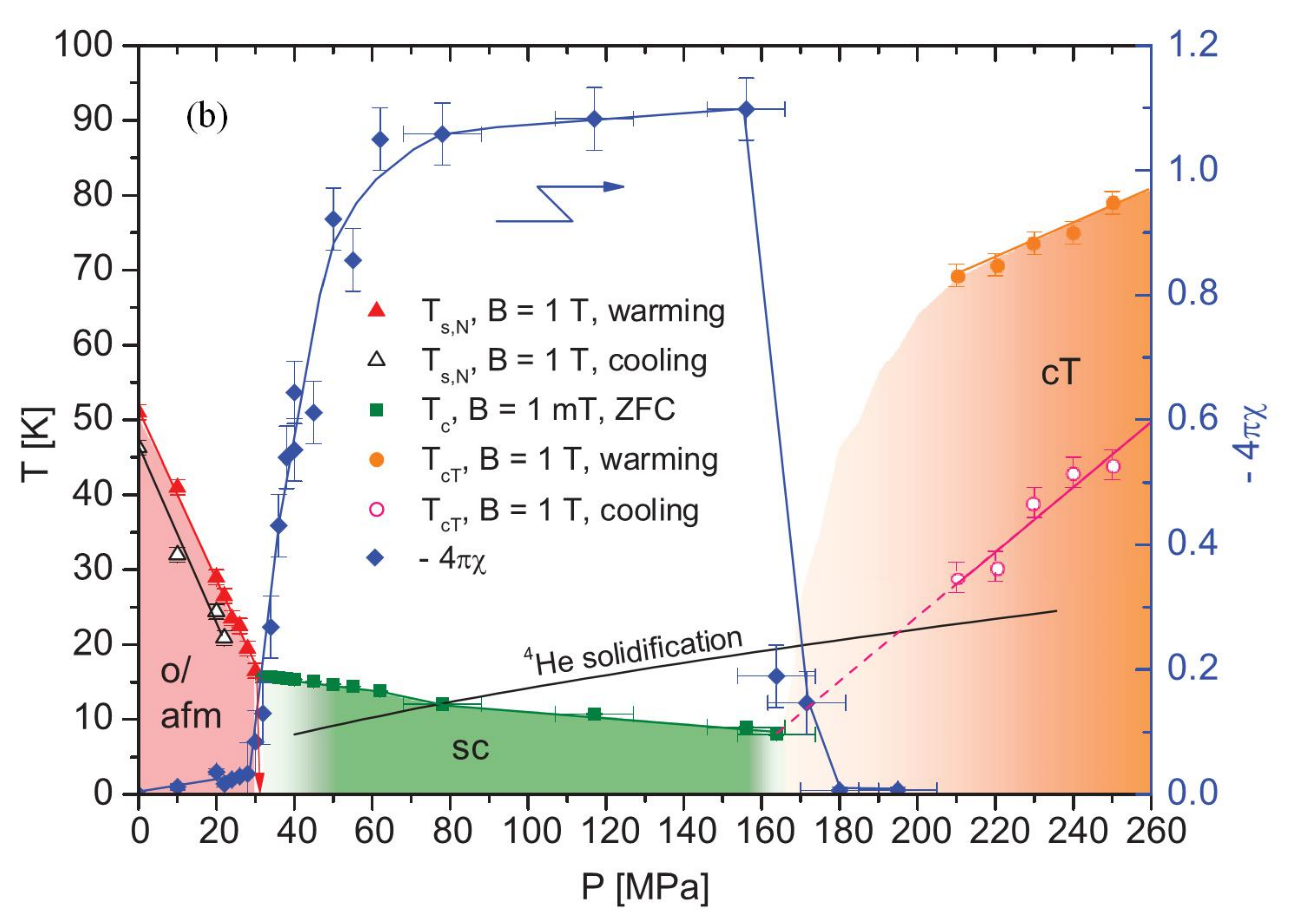}
\end{center}
\caption{(Color online) (a)  Three dimensional phase diagram with substitution level, $x$, annealing/quenching temperature, $T_{A/Q}$, and transition temperature, $T$, as three axes. Red (AFM/ORTH), green (SC/PM/T) and blue (CT) spheres represent data. Transparent, colored surfaces are guides to the eyes. (after Refs. \cite{ran12a,ran14a}); (b) $T - P$  phase diagram of  Ca(Fe$_{1-x}$Co$_x$)$_2$As$_2$ with $x = 0.028/T_{anneal} = 350^\circ$ C inferred from
magnetic susceptibility data \cite{gat12a}.} \label{F7}
\end{figure}

\clearpage

\begin{figure}[htbp]
\begin{center}
\includegraphics[angle=0,width=100mm]{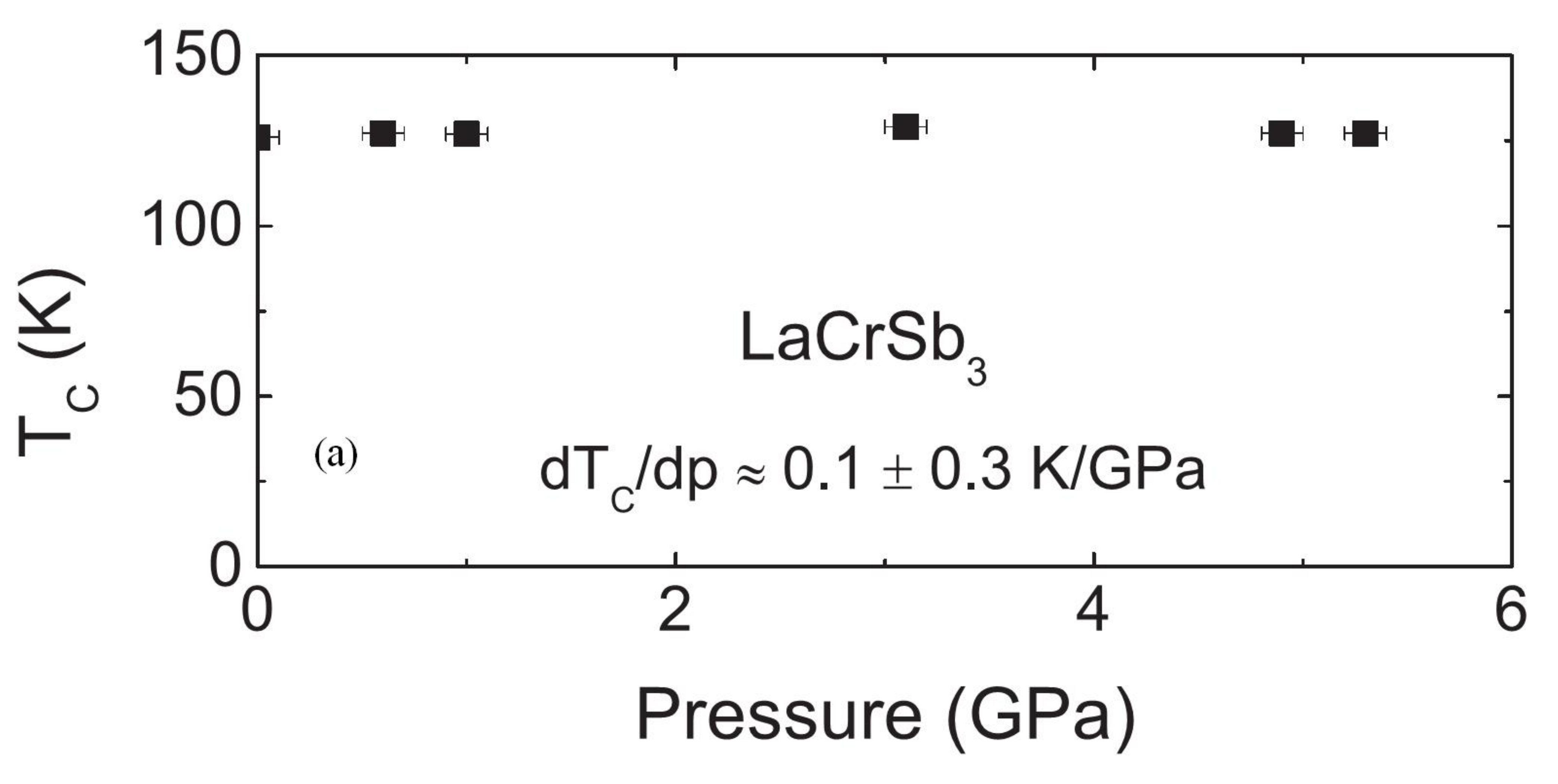}
\includegraphics[angle=0,width=100mm]{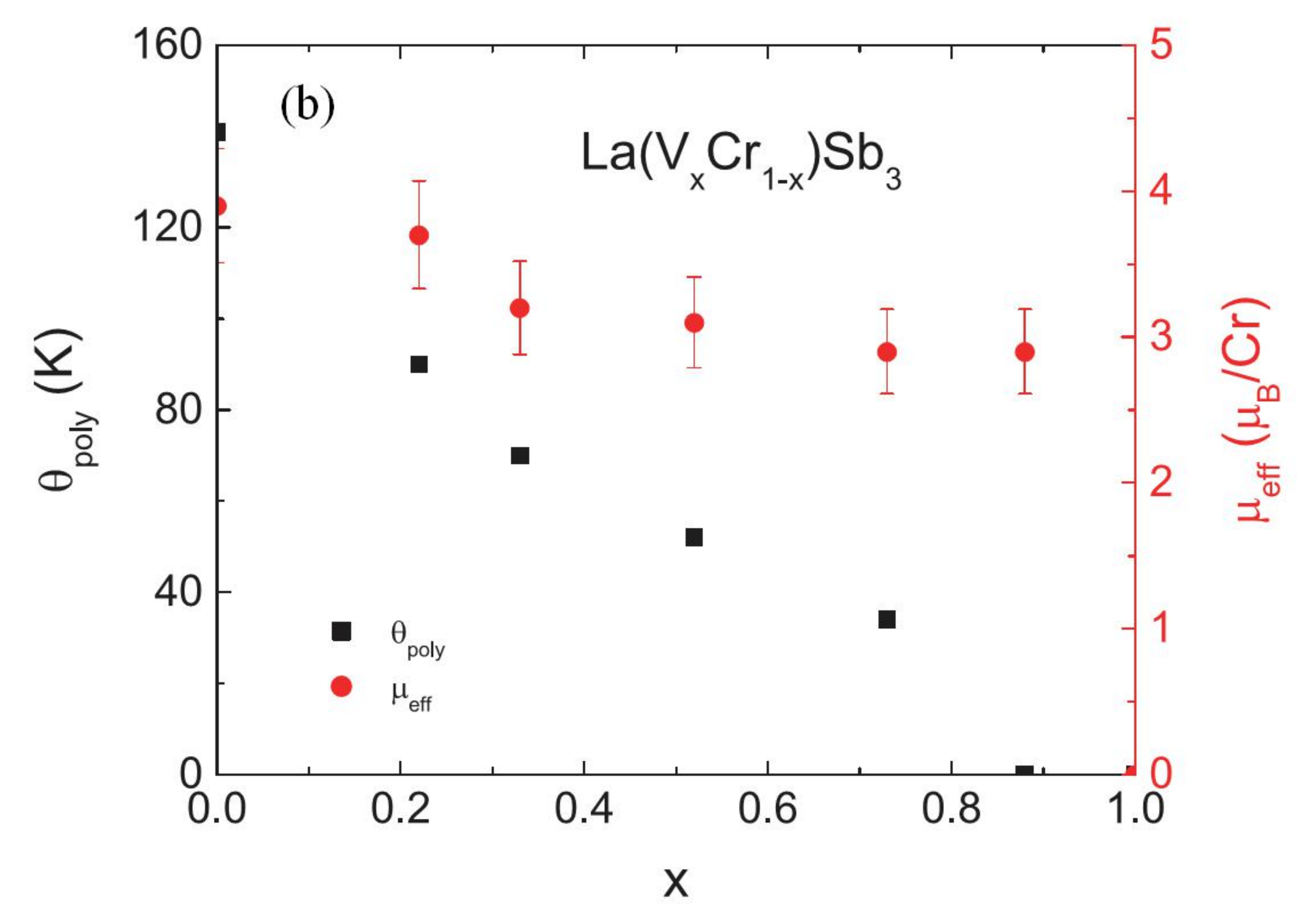}
\includegraphics[angle=0,width=90mm]{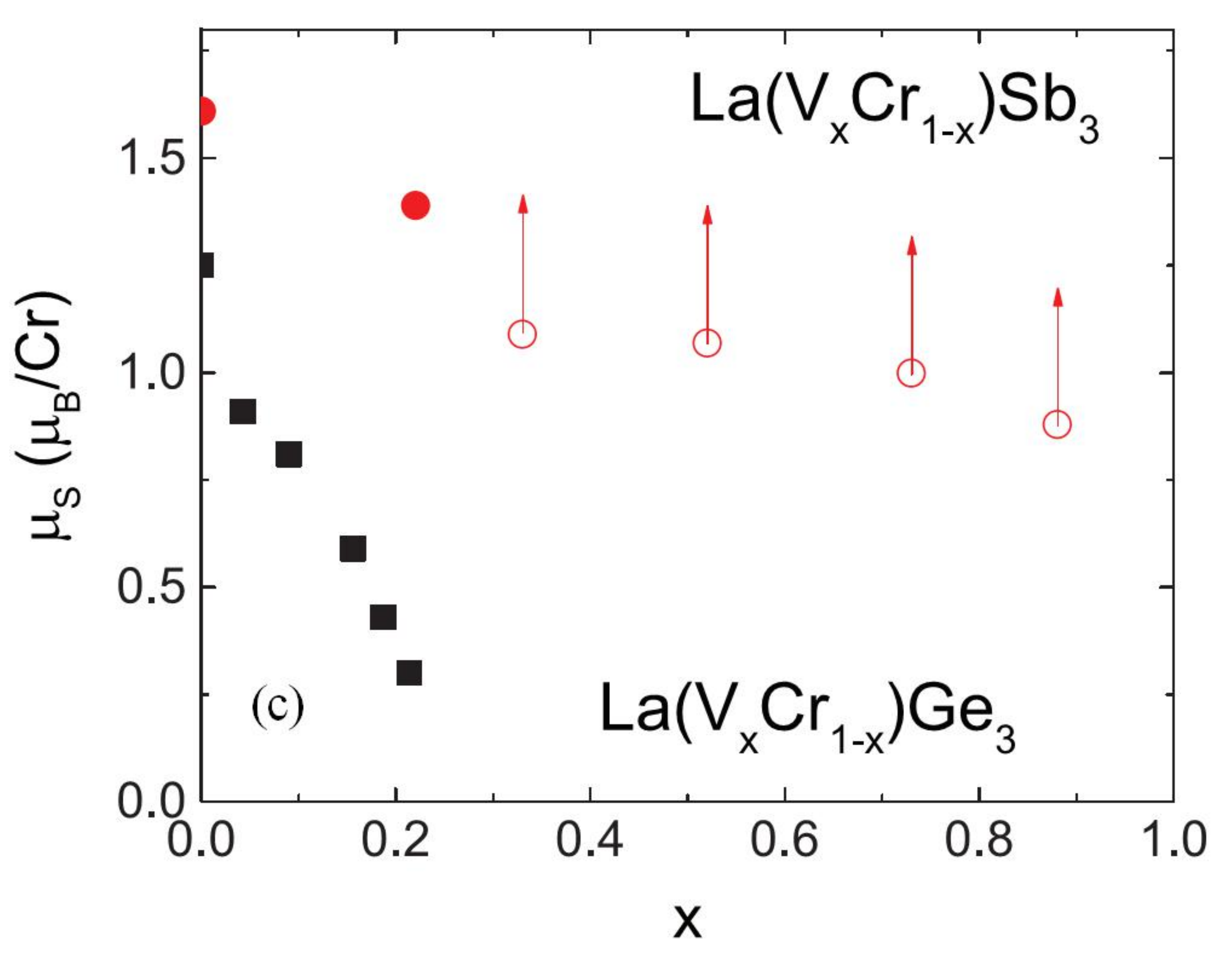}
\end{center}
\caption{(Color online) (a) Pressure dependence of $T_C$ for LaCrSb$_3$; (b) The Curie-Weiss temperature $\theta_{poly}$ and the effective moment $\mu_{eff}$ per Cr as a function of $x$ for  La(V$_x$Cr$_{1-x}$)Sb$_3$; (c) The saturated moment $\mu_S$ as a function of $x$ for the La(V$_x$Cr$_{1-x}$)Ge$_3$ and La(V$_x$Cr$_{1-x}$)Sb$_3$ series \cite{lin14a}.} \label{F8}
\end{figure}

\clearpage

\begin{figure}[htbp]
\begin{center}
\includegraphics[angle=0,width=120mm]{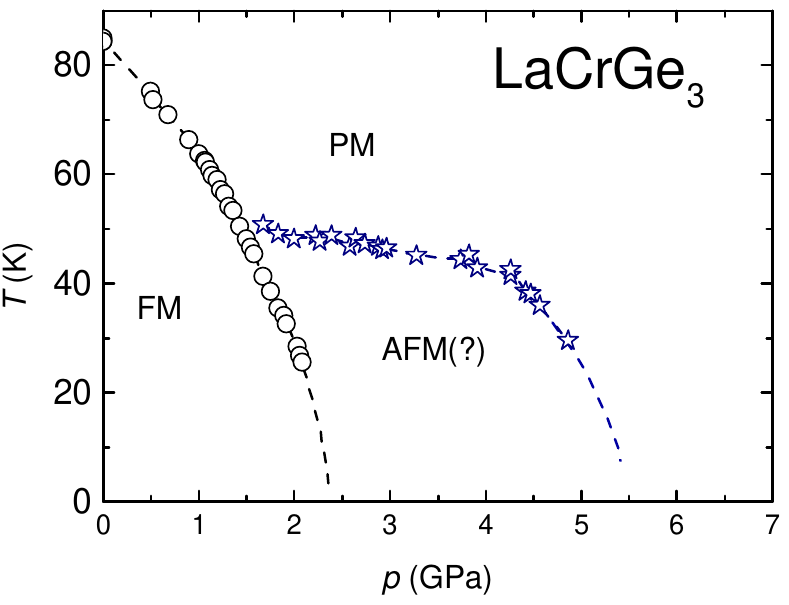}
\includegraphics[angle=0,width=130mm]{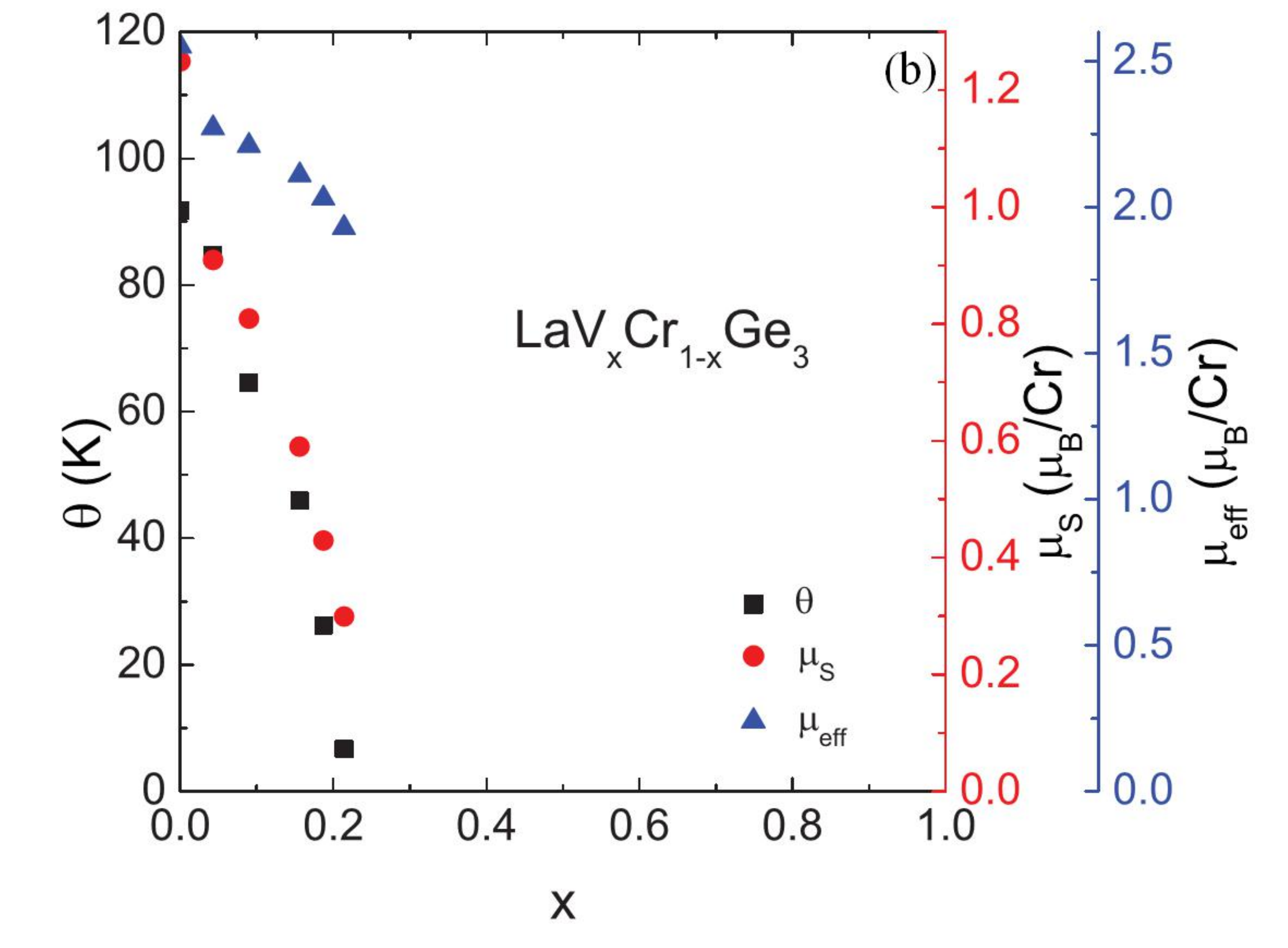}
\end{center}
\caption{(Color online) (a)  Schematic $T - p$ phase diagram for LaCrGe$_3$ from resistivity measurements (after Ref. \cite{tau16a,kal16a}); (b) the Curie-Weiss temperature $\theta$, saturated moment $\mu_S$ along the $c$ axis, and effective moment $\mu_{eff}$ per Cr as a function of $x$ for La(V$_x$Cr$_{1-x}$)Ge$_3$  \cite{lin13a}.} \label{F9}
\end{figure}

\end{document}